\documentclass[12pt]{iopart}

\usepackage{aas_macros}
\usepackage{psfrag} 

\usepackage{graphicx}
\usepackage{txfonts}
\usepackage{bm}
\usepackage{soul}
\usepackage[svgnames]{xcolor}

\newcommand{\vv}{\mathcal{V}}

\newcommand{\LL}{\mathcal{L}}

\newcommand{\nn}{\mathfrak{N}}

\newcommand{\deltx}{\Delta}
\newcommand{\deltt}{\Delta}

\newcommand{\timex}{ }

\newcommand{\jcop}{J.~Comput.~Phys.}

\newcommand{\eqref}[1]{Eq.\,(\ref{#1})}

\newcommand{\figref}[1]{Fig.\,\ref{#1}}
\newcommand{\tabref}[1]{Tab.\,\ref{#1}}
\newcommand{\secref}[1]{Sec.\,\ref{#1}}

\begin{document}

\title[Numerical viscosity in the KH instability]{Numerical viscosity
  in simulations of the two-dimensional Kelvin-Helmholtz instability}


\author{M.~Obergaulinger${}^{1}$, M.\'A.~Aloy${}^{2}$}
\address{
  ${}^1$ Institut f{\"u}r Kernphysik, Theoriezentrum, S$2|11$, Schlo{\ss}gartenstr.~2, 64289 Darmstadt, Germany
  \\
  ${}^2$ Departament d'Astronomia i Astrof{\'i}sica, Universitat de Val{\`e}ncia, 
  Edifici d{\'{}}Investigaci{\'o} Jeroni Munyoz, C/ Dr.~Moliner, 50,
  E-46100 Burjassot (Val{\`e}ncia), Spain
}
\ead{mobergaulinger@theorie.ikp.physik.tu-darmstadt.de}

\begin{abstract}
  The Kelvin-Helmholtz instability serves as a simple, well-defined
  setup for assessing the accuracy of different numerical methods for
  solving the equations of hydrodynamics.  We use it to extend our
  previous analysis of the convergence and the numerical dissipation
  in models of the propagation of waves and in the tearing-mode
  instability in magnetohydrodynamic models.  To this end, we perform
  two-dimensional simulations with and without explicit physical
  viscosity at different resolutions.  A comparison of the growth of
  the modes excited by our initial perturbations allows us to estimate
  the effective numerical viscosity of two spatial reconstruction
  schemes (fifth-order monotonicity preserving and second-order
  piecewise linear schemes).
\end{abstract}

\section{Introduction}
\label{Sek:Intro}

The finite resolution of numerical simulations of
(magneto-)hydrodynamical ((M)HD) systems inevitably causes numerical
errors.  Convergence requires that these errors approach zero for
increasing grid resolution.  With this limit being never practically
attainable, a quantitative characterisation of the errors can be a
valuable tool in assessing the results of a simulation.  A useful way
of doing so relies on an analogy to physical diffusion and
dissipation, expressing the errors in terms of numerical shear and
bulk viscosities and, for non-zero magnetic fields, a numerical
resistivity.

All of these quantities may depend in a complex way on the numerical
scheme and the spatial and temporal resolution as well as on the
specific physical system simulated.  A possible form of these
dependencies was suggested by
\cite{Rembiasz_et_al__2017__apjs__OntheMeasurementsofNumericalViscosityandResistivityinEulerianMHDCodes}.
They proposed that, e.g., the numerical shear viscosity, $\nu_{\ast}$,
is the sum of two contributions due to the temporal and the spatial
discretisation, which are proportional to problem-dependent
characteristic velocity, $\vv$, and length scales, $\LL$, and to a
power of the time step, $\Delta t$, and grid width, $\Delta x$,
respectively:
\begin{equation}
  \label{Gl:nu*}
  \nu_{\ast} = \nn_{\nu}^{\deltx x} \timex \vv \timex \LL \timex 
  \left( \frac{\deltx x}{\LL}\right)^{r} + 
  \nn_{\nu}^{\deltx t} \timex \vv \timex \LL \timex 
  \left( \frac{\vv \deltt t }{\LL}\right)^{q}.
\end{equation}
The normalisation coefficients, $\nn_{\nu}^{\deltx x}$ and
$\nn_{\nu}^{\deltx t}$, and the exponents $r,q$ depend on the
numerical scheme.  Put to the test in a series of one- and
two-dimensional problems with known solutions such as the propagation
of various MHD waves or the resistive tearing-mode instability, the
ansatz proved to be a good description of the errors of the Eulerian
code used in the study.  The spatial reconstruction of very high order
(a monotonicity preserving scheme based on polynomials of up to ninth
order; \cite{Suresh_Huynh__1997__JCP__MP-schemes}) was found to yield
correspondingly high orders of convergence in the numerical
resistivities and viscosities expressed by exponents of up to $r
\approx 9$.  

Our goal is to study the dependence of numerical errors for the same
code in another system, viz.~the hydrodynamic Kelvin-Helmholtz (KH)
instability (KHI).  The present study represents a first steps towards a
full description similar to our previous work.  The practical benefit
of our ansatz is somewhat restricted by an ignorance of $\vv$ and, in
particular, $\LL$, which, as shown for the tearing modes, can depend
in a quite intricate manner on both physical and numerical parameters.
Hence, we will not aim at an identification of the unknown coefficients
in \eqref{Gl:nu*} but restrict ourselves to a quantitative
determination of the overall rates of numerical dissipation.

The KHI of a shear layer has been used extensively as a test for
numerical codes.  Among the various setups proposed for this purpose,
we use the one devised by
\cite{McNally_et_al__2012__apjs__AWell-posedKelvin-HelmholtzInstabilityTestandComparison}
who found numerical convergence during the linear phase of growth of
the instability.  After the growth saturates, secondary instabilities
develop and make the dependence of the evolution to the numerical
resolution more complicated
\cite{Lecoanet_et_al__2016__mnras__Avalidatednon-linearKelvin-Helmholtzbenchmarkfornumericalhydrodynamics}.
Therefore, we defer a more thorough investigation of this stage to a
later study.

This article is organised as follows: \secref{Sek:NumIni} describes
the numerical methods used for the simulations and the initial data,
\secref{Sek:Res} presents the results, and \ref{Sek:Concl} summarises
the main outcomes.

\section{Numerical methods and initial conditions}
\label{Sek:NumIni}

The numerical code used here is the same as in
\cite{Rembiasz_et_al__2017__apjs__OntheMeasurementsofNumericalViscosityandResistivityinEulerianMHDCodes}
(see also \cite{Obergaulinger__2008__PhD__RMHD}).  It is based on a
Eulerian finite-volume discretisation of the MHD equations in the
constrained-transport framework \cite{Evans_Hawley__1998__ApJ__CTM}
and employs high-order reconstruction methods, Runge-Kutta time
integrators of up to fourth order, and approximate Riemann solvers.
In the simulations reported here, we compare spatial reconstruction
using the fifth-order monotonicity preserving method (MP5) of
\cite{Suresh_Huynh__1997__JCP__MP-schemes} to a second-order piecewise
linear scheme (PLM).  While we restrict ourselves here to viscous
Newtonian hydrodynamics, we note that a wide range of additional
physics is implemented such as self-gravity and neutrino transport
\cite{Just_et_al__2015__mnras__Anewmultidimensionalenergy-dependenttwo-momenttransportcodeforneutrino-hydrodynamics}
as well as a wider range of equations of state
(\cite{Aloy_et_al__2019__mnras__Neutronstarcollapseandgravitationalwaveswithanon-convexequationofstate}
and references therein).

Following
\cite{McNally_et_al__2012__apjs__AWell-posedKelvin-HelmholtzInstabilityTestandComparison},
our setup consists of a quadratic box ($0 \le x \le 1$, $0 \le y \le
1$) with $n^x = n^y = n$ zones per direction.  Boundary conditions in
$x$ and $y$-directions are periodic.  The grid contains two smooth
shear layers with a gradient in the $x$-component of the velocity,
$\vec v$, and a contrast of the density, $\rho$, given by
\begin{eqnarray}
  \label{Gl:vx-initt}
  v^x & = & 
  \left\{
    \begin{array}{rl}
      U_1 - U_m \exp{ \frac{y - 1/4}{L}} & \mathrm{for } ~ 0 \leq y \le 1/4, \\
      U_2 + U_m \exp{ \frac{y + 1/4}{L}} & \mathrm{for } ~ 1/4 \leq y \le 1/2, \\
      U_2 + U_m \exp{ \frac{y - 3/4}{L}} & \mathrm{for } ~ 1/2 \leq y \le 3/4, \\
      U_1 - U_m \exp{ \frac{y + 3/4}{L}} & \mathrm{for } ~ 3/4 \leq y \le 1,
    \end{array}
  \right.
  \\
  \rho  & = & 
  \left\{
    \begin{array}{rl}
    \rho_1 - \rho_m \exp{ \frac{y - 1/4}{L}} & \mathrm{for } ~ 0 \leq y \le 1/4, \\
    \rho_2 + \rho_m \exp{ \frac{y + 1/4}{L}} & \mathrm{for } ~ 1/4 \leq y \le 1/2, \\
    \rho_2 + \rho_m \exp{ \frac{y - 3/4}{L}} & \mathrm{for } ~ 1/2 \leq y \le 3/4, \\
    \rho_1 - \rho_m \exp{ \frac{y + 3/4}{L}} & \mathrm{for } ~ 3/4 \leq y \le 1,
    \end{array}
  \right.
\end{eqnarray}
with $U_{1,2} = \pm 0.5$, $\rho_1 = 1, \rho_2 = 2$, $._m = (._1 -
._2)/2$, and $L = 0.025$.  We add the following perturbation in the $y$-component of the
velocity:
\begin{equation}
  \label{Gl:vy-init}
  v^y = 0.01 \sin ( 4 \pi x).
\end{equation}
The gas obeys an ideal-gas EOS with $\gamma = 5/3$ and initially has a uniform
pressure of $P = 2.5$.

Several series of simulations with resolutions between $n = 128$ and
$n = 4096$ cells were run.  We used shear viscosities between $\nu =
0$ (ideal HD) and $\nu = 10^{-3}$.  Most runs employed the MP5
spatial reconstruction and a third-order Runge-Kutta time stepping,
but we added models with PLM reconstruction to test
the influence of the numerical methods.  We will refer to the
non-viscous simulation with the highest resolution, $n = 4096$, as the
reference run.  See \tabref{Tab:runs} for an overview of our simulations.

\begin{table}
  \centering
  \begin{tabular}{|r|rrrrrrrrr|r|r}
    \hline
    & \multicolumn{9}{c|}{$\nu$ (MP5)} & PLM
    \\ 
    $n$ & $10^{-3}$ & $10^{-4}$ & $3 \times 10^{-5}$ & $10^{-5}$ & $3
                                                                 \times
                                                                 10^{-6}$
                                                               &
                                                                 $10^{-6}$
                                                               &
                                                                 $10^{-7}$ & $10^{-8}$ & 0 & 0
    \\
    \hline
    128 & $\surd$ & $\surd$ & & $\surd$ & & $\surd$ & $\surd$ & & $\surd$ & $\surd$ 
    \\
    256 & $\surd$ & $\surd$ & $\surd $& $\surd$ & $\surd$ & $\surd$ & $\surd$ & & $\surd$ & $\surd$ 
    \\
    512 & $\surd$ & $\surd$ & $\surd $& $\surd$ & $\surd$ & $\surd$ & $\surd$ & & $\surd$ & $\surd$ 
    \\
    1024 & $\surd$ & $\surd$ & & $\surd$ & & $\surd$ & $\surd$ &$\surd $ & $\surd$ & $\surd$ 
    \\
    2048 &  & $\surd$ & & $\surd$ & & $\surd$ & $\surd$ &$\surd $ & $\surd$ & $\surd$ 
    \\
    4096 &  & & & &&&& & $\surd$ &
    \\
    \hline
  \end{tabular}
  \caption{
    List of runs.  For each resolution, $n$, we mark the models we
    performed.  Runs with MP5 reconstruction have used a range of
    explicit viscosities as indicated in the middle block of the
    table, whereas runs with PLM reconstruction were all performed in
    ideal HD.  The simulation with $n = 4096$ is the reference run.
  }
  \label{Tab:runs}
\end{table}

We run the simulations up to a final time of $t_{\mathrm{f}} = 3$,
i.e., past the end of the linear growth phase of the KHI and into the
transition to the non-linear saturated phase.  We analyse the growth
using the amplitude, $M$, of the initially excited mode introduced by
\cite{McNally_et_al__2012__apjs__AWell-posedKelvin-HelmholtzInstabilityTestandComparison}:
\begin{eqnarray}
  \label{Gl:di}
  d & = &
  \left\{
    \begin{array}{rl}
      \exp (-4 \pi |y - 0.25|) & \mathrm{for ~} y \le 1/2, \\
      \exp (-4 \pi |y - 0.75|) & \mathrm{for ~} y \ge 1/2,
    \end{array}
  \right.
  \\
  s & = & v^y \sin (4 \pi x) \, d,
  \\
  c & = & v^y \cos (4 \pi x) \, d,
  \\
  M & = &
  2 \sqrt{
    \left(\frac{\int_V \mathrm{d} V s}{\int_V \mathrm{d} V d}\right)^2
    +
    \left(\frac{\int_V \mathrm{d} V c}{\int_V \mathrm{d} V d}\right)^2
  }.
\end{eqnarray}
The integrals in the last equation extend over the numerical domain.
To quantify the numerical errors of a specific run, we define the
deviation of its mode amplitude from that of the reference run and the
corresponding time integrals,
\begin{eqnarray}
  \label{Gl:Dt}
  D & = & M - M_{\mathrm{ref}},
  \\ 
  \label{Gl:Dintt15}
  E_{1.5} & = & \int_{0}^{1.5} \mathrm{d} t \, D (t),
  \\ 
  \label{Gl:Dintt3}
  E_{3} & = & \int_{0}^{3} \mathrm{d} t \, D (t).
\end{eqnarray}

\section{Results}
\label{Sek:Res}

\begin{figure}
  \centering
  \includegraphics[width=\linewidth]{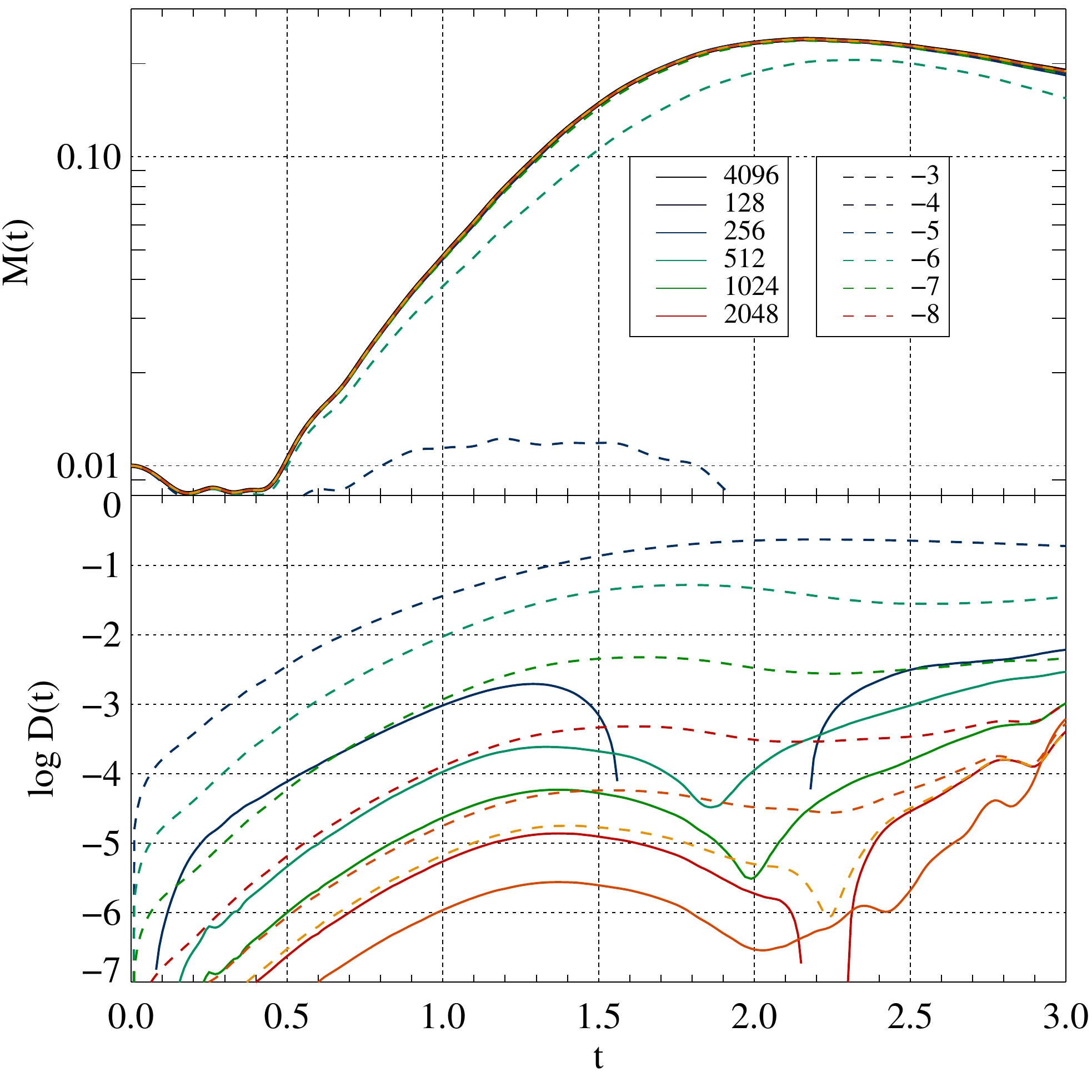}
  \caption{
    Time evolution of the velocity mode amplitude, $M$ (top), of several
    simulations and their deviation from the reference run, $D$
    (bottom).  Solid lines represent models without explicit viscosity
    with colours distinguishing between different grid resolutions,
    while dashed lines show runs with the same grid of $n = 1024$ and
    different shear viscosities from $\nu = 10^{-3}$ to $\nu =
    10^{-8}$ (the legend shows the logarithm of $\nu$).  
  }
  \label{Fig:khMt}
\end{figure}

\begin{figure*}
  \centering
  \includegraphics[width=0.32\linewidth]{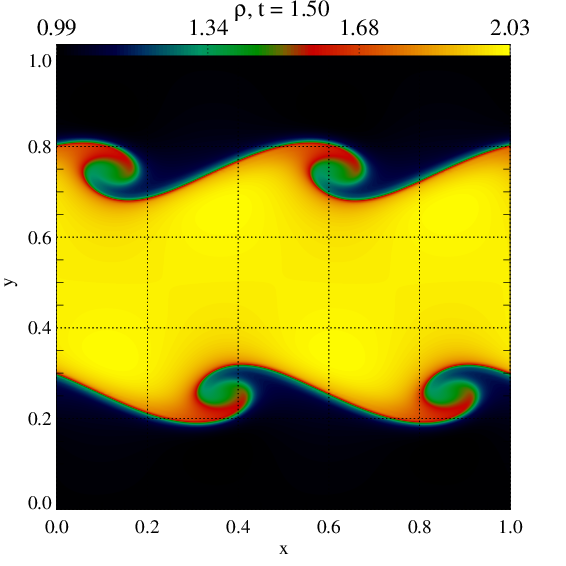}
  \includegraphics[width=0.32\linewidth]{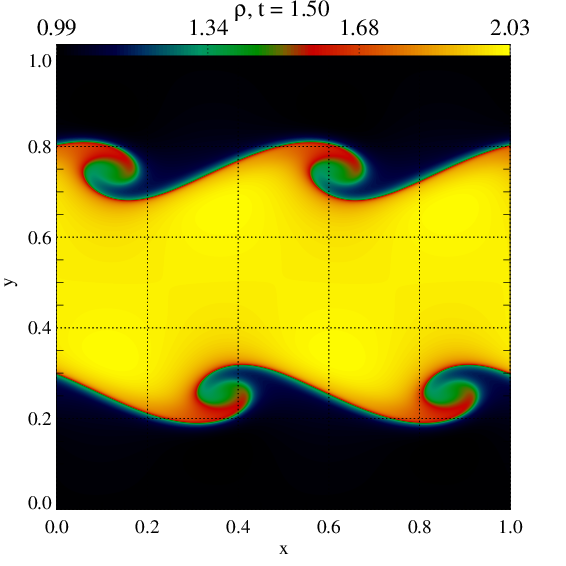}
  \includegraphics[width=0.32\linewidth]{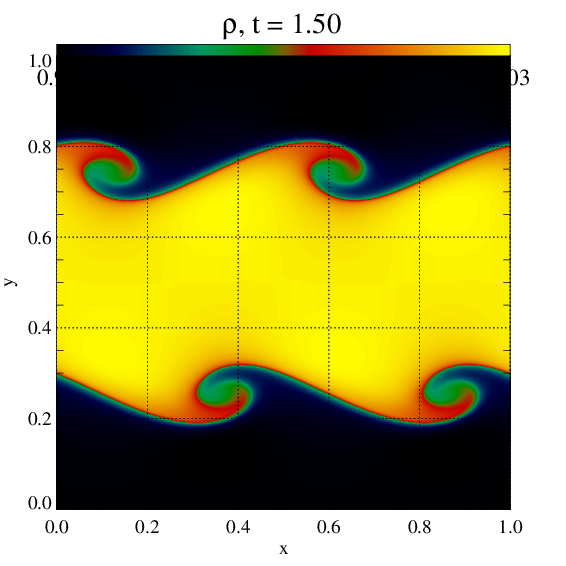}
  \includegraphics[width=0.32\linewidth]{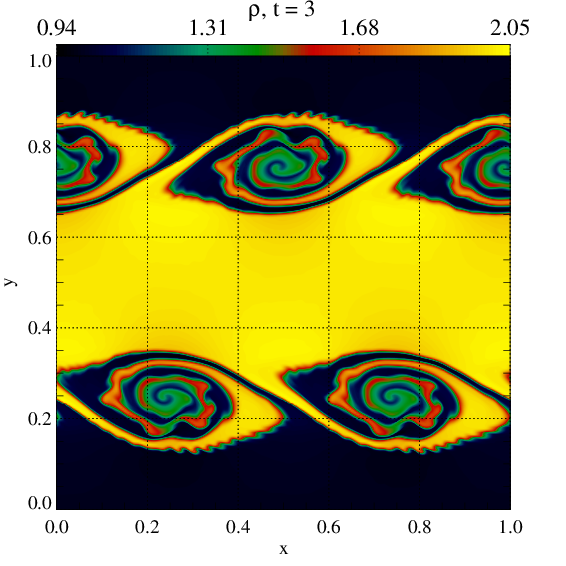}
  \includegraphics[width=0.32\linewidth]{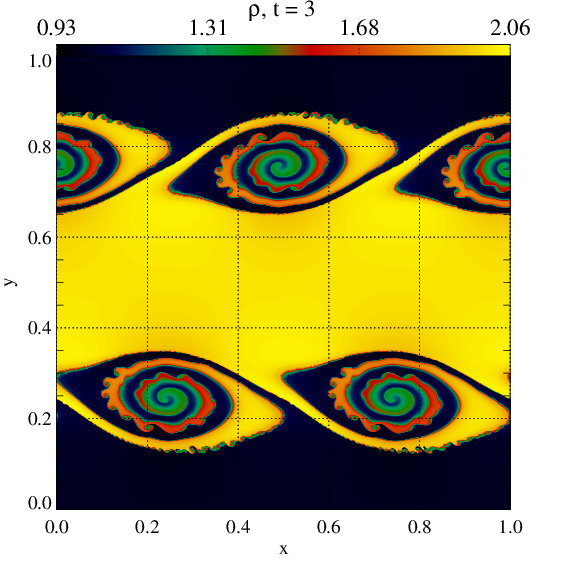}
  \includegraphics[width=0.32\linewidth]{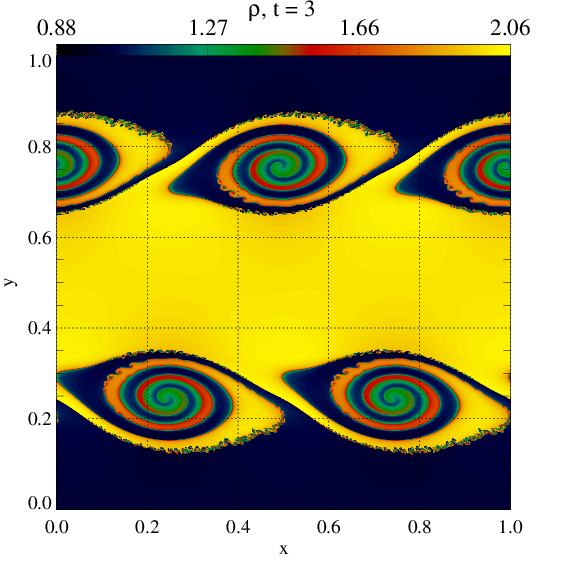}
  \caption{
    Density maps at $t = 1.5$ (top) and $t = 3$ (bottom) of the
    non-viscous runs with, from left to right, $n = 256, 1024, 4096$.
  }
  \label{Fig:KH2d}
\end{figure*}

The top panel of \figref{Fig:khMt} shows the time evolution of the
mode amplitude $M$ of several models without (solid lines) and with
(dashed lines) explicit viscosity.  The reference run (solid black
line) and all simulations with a viscosity $\nu \le 10^{-5}$ evolve in
a very similar way with only small quantitative differences.  After a
transient decrease, $M$ starts to rise at $t \approx 0.4$.  The
increase proceeds at first exponentially with time.  It slows down
after $t \approx 1.3$ and reaches a maximum at $t \approx 2.1$.
During this phase, the flow develops the characteristic KH vortices
(see the density maps at $t = 1.5$ in the top panels of
\figref{Fig:KH2d}) After the saturation of growth, $M$ gradually
declines.  The vortices continue to wind up and secondary instablities
grow on the thin filaments around their centres (for $t = 3$, see
bottom panels of \figref{Fig:KH2d}).  Similarly to
\cite{McNally_et_al__2012__apjs__AWell-posedKelvin-HelmholtzInstabilityTestandComparison}
and
\cite{Lecoanet_et_al__2016__mnras__Avalidatednon-linearKelvin-Helmholtzbenchmarkfornumericalhydrodynamics}
(cf.~their Fig.~10), we can distinguish between instabilities of the
outer filaments and those of the inner core.  The latter develop near
the centres of the vortices, while the former appear near their edges.
As found by 
\cite{McNally_et_al__2012__apjs__AWell-posedKelvin-HelmholtzInstabilityTestandComparison}
and
\cite{Lecoanet_et_al__2016__mnras__Avalidatednon-linearKelvin-Helmholtzbenchmarkfornumericalhydrodynamics},
the outer filament instabilities are present at all resolutions,
whereas the inner core instabilities grow slower at higher resolution
to finally disappear for $n = 4096$.

The growth of the KHI is noticeably smaller for very high physical
viscosity of $\nu = 10^{-4}$ and completely absent for
$\nu = 10^{-3}$.  In these cases, the diffusion times for smearing out
the shear layers is roughly
$\tau_{\mathrm{diff}} \sim \frac{L^2}{\nu} = 6.25
\frac{10^{-4}}{\nu}$, i.e., of a similar order of magnitude as the
growth time scale of the KHI.  Diffusion decreases the gradient of
$v^x$ across the shear layer, providing an additional decrease of the
development of the KHI ($\nu = 10^{-4}$) or even completely
suppressing it ($\nu = 10^{-3}$).  Evidence for this explanation was
found in auxiliary simulations in which the viscosity only operates on
the difference of the velocity from its initial state.  With diffusion
of the shear layer thus eliminated, the additional suppression of the
KHI is absent.

Since, except for the highest viscosities, all runs display a very
similar evolution during the linear growth phase, we can use this
stage without further complications in a convergence study.  To this
end, we examine the deviation from the reference, $D$, run until $t =
1.5$.  As shown in the bottom panel of \figref{Fig:khMt}, $D$, and
hence ${E_{1.5}}$, approach $0$ as $n \to 4096$.

\begin{figure}
  \centering
  \includegraphics[width=0.49\linewidth]{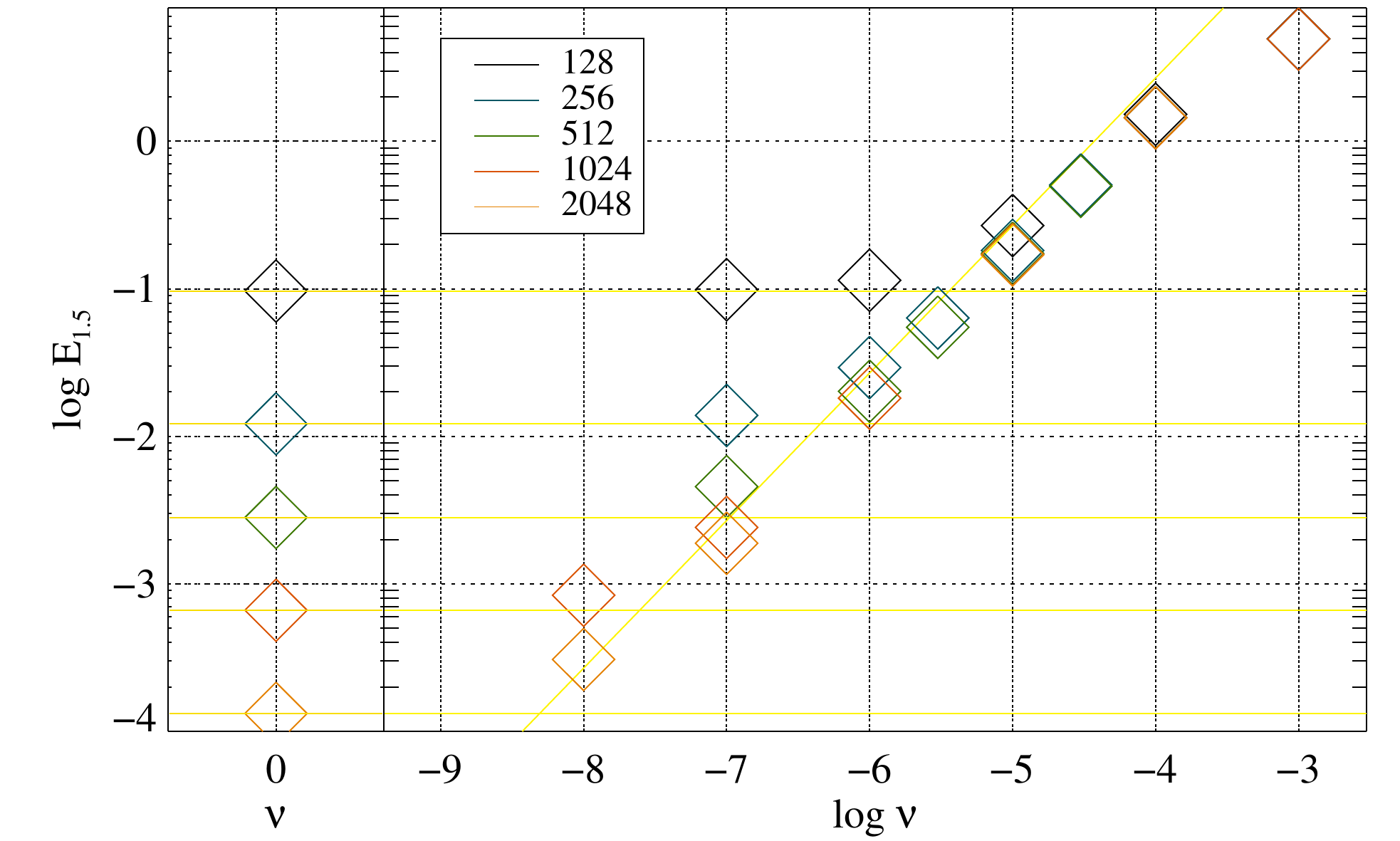}
  \includegraphics[width=0.49\linewidth]{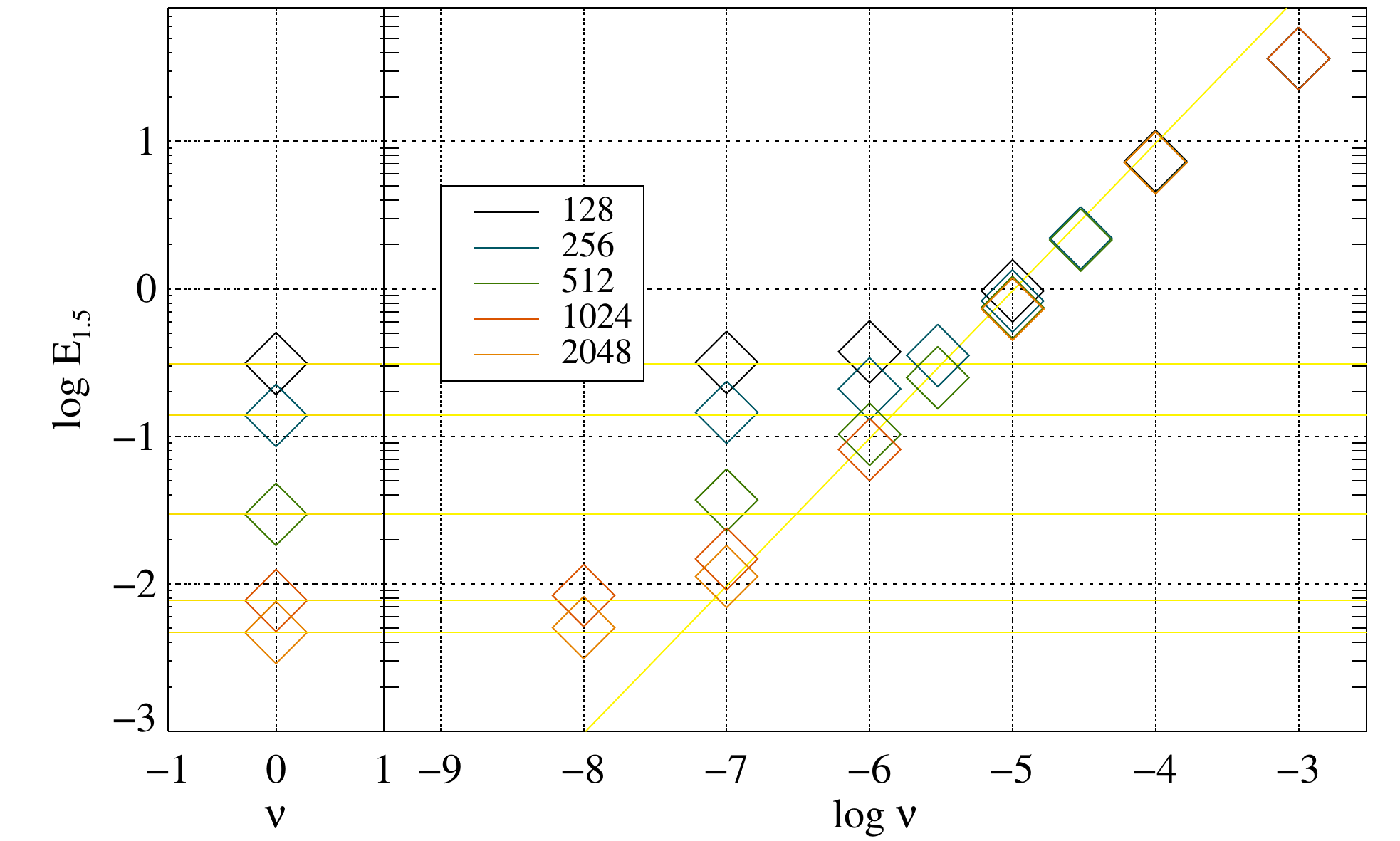}
  \caption{
    Left:
    Time-integrated deviation from the reference run, $E_{1.5}$, for
    simulation series with different resolutions (colours) as
    functions of viscosity (all with MP5 reconstruction).  The horizontal yellow lines show the
    value of $E_{1.5}$ in the simulations in ideal HD.  The inclined
    yellow line displays a linear relation $E_{1.5} \propto \nu$.
    Right:
    The same, but for $E_3$.
  }
  \label{Fig:Ereso}
\end{figure}

The left panel of \figref{Fig:Ereso}, displaying $E_{1.5}$ as a function of viscosity
for runs with different resolutions, displays two limiting behaviours.
At high viscosity, $E_{1.5}$ follows a linear relation, $E_{1.5}
\propto \nu$ that is common for all grids.  The deviation of
simulations at $\nu = 10^{-3,4}$ from this trend owes itself to the
smearing out of the shear layer due to rapid diffusion.  If $\nu$ is
decreased below a resolution-dependent threshold, $E_{1.5}$ levels off
and approaches the value of the simulation in ideal HD, for which the
diffusion is entirely due to numerical errors.

\begin{figure}
  \centering
  \includegraphics[width=0.49\linewidth]{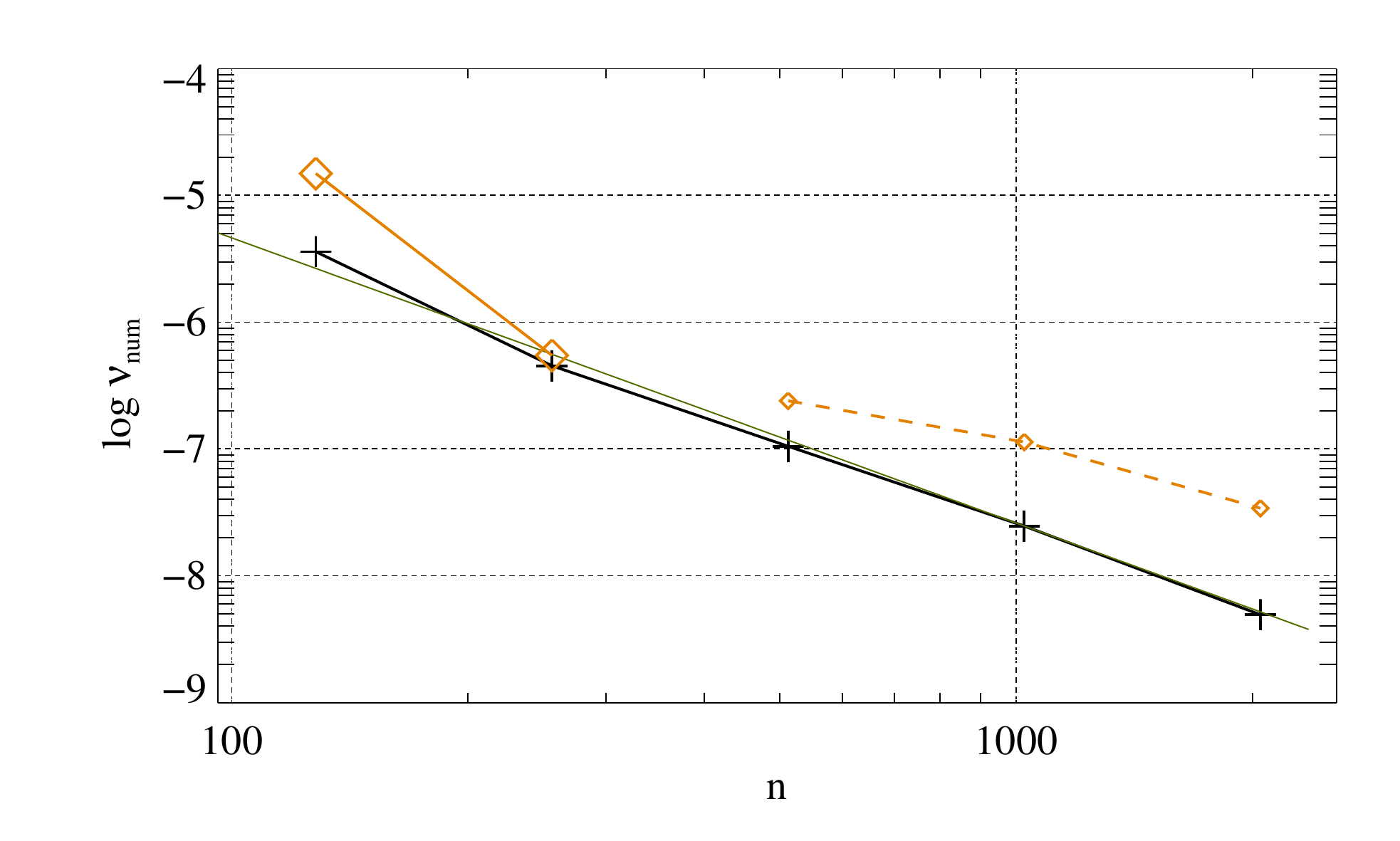}
  \includegraphics[width=0.49\linewidth]{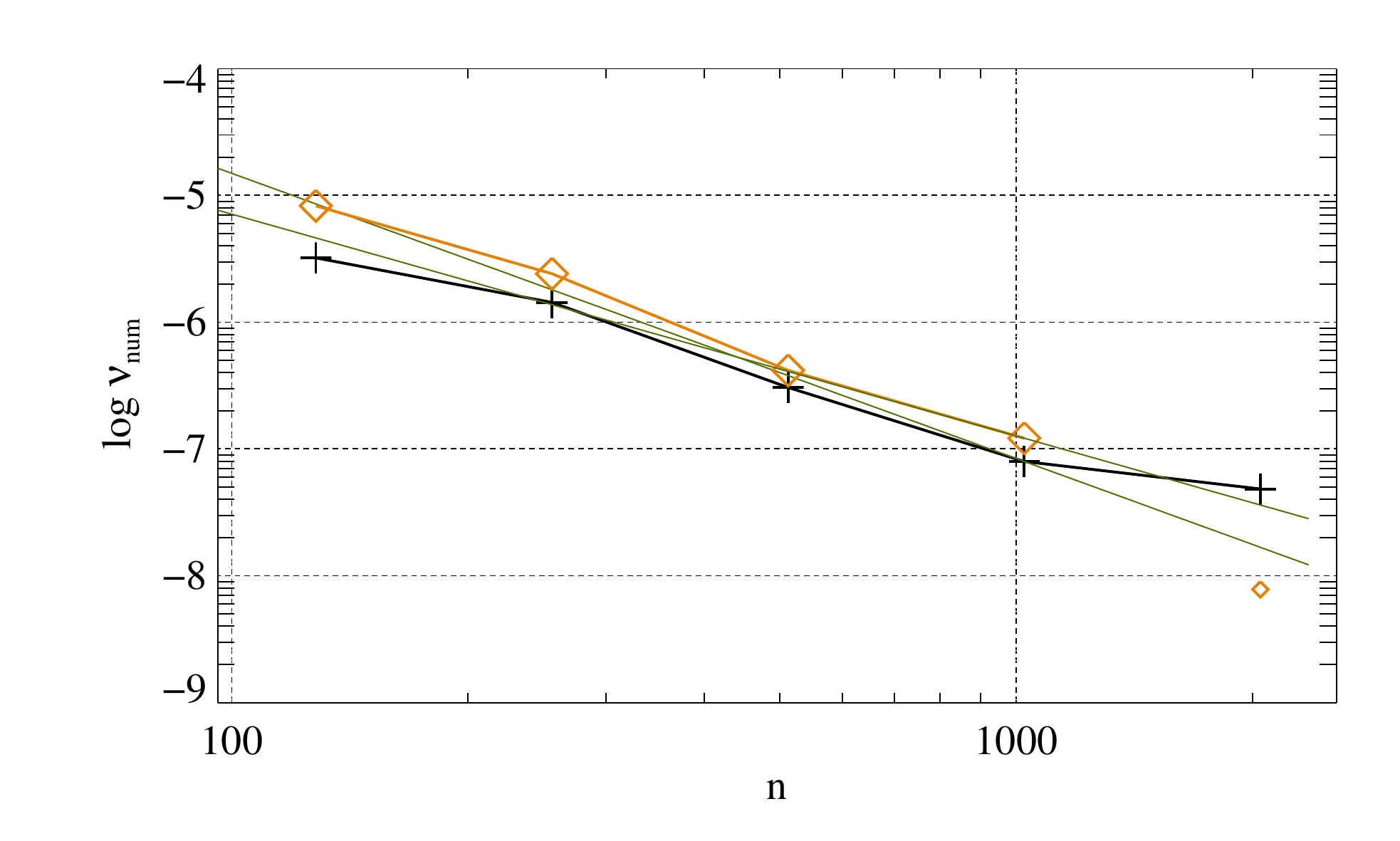}
  \caption{
    Left:
    The effective numerical viscosity of ideal-HD runs as a function
    of grid resolution.  Black and orange symbols correspond to runs
    with MP5 and PLM reconstruction, respectively.  For the latter
    group of runs, $E_{1.5}$ changes sign between $n = 256$ and $n =
    512$.  The two domains with different signs are distinguished by
    line styles.  The green line represents the power law
    approximation \eqref{Gl:nunum}.
    Right: same for $E_3$.
  }
  \label{Fig:numun}
\end{figure}

We can surmise that this numerical finding represents a physical
result, i.e., that dissipation in the growth phase of the KHI operates
in such a way that $E_{1.5}$ increases linearly with the viscosity.
In our setup, the scaling is
\begin{equation}
  \label{Gl:E15nu}
  E_{1.5} \approx 0.27 \frac{\nu}{10^{-5}}.
\end{equation}
The close similarity between a simulation in ideal HD and one oln a
finer grid, but with physical viscosity, serves as an indication that
the numerical errors indeed behave like a physical viscosity, which
allows us to apply the scaling relation, \eqref{Gl:E15nu}, to
determine the effective numerical viscosity of the code.  The results,
shown in  \figref{Fig:numun}, suggest for runs with
MP5 reconstruction a power-law dependence of the numerical viscosity,
$\nu_{\mathrm{num}}$, on the resolution:
\begin{equation}
  \label{Gl:nunum}
  \nu_{\mathrm{num}} = 2.5 \times 10^{-8} \left(\frac{n}{1024}\right)^{-2.25}.
\end{equation}
The numerical errors of runs with second-order TVD piecewise-linear
(PLM) reconstruction depend on resolution in a less straightforward
way (orange line).  For the finest grids ($n \ge 512$), we find a
slightly shallower scaling relation $\nu_{\mathrm{num}} \propto n ^
{-1.75}$ at about one order of magnitude above the MP5 runs.  The
deviations from the scaling at the coarsest grids are stronger than
for runs with MP5, even going so far as to change sign between $n =
256$ and $n = 512$.

We note that other elements of the numerical scheme such as the time
integrators, CFL number, and Riemann solvers have a smaller influence
on $E_{1.5}$ then the reconstruction method.

Applying the same analysis to the entire simulation time (for $E_3$,
see the right panel of \figref{Fig:Ereso} and for the resulting
numerical viscosity the right panel of \figref{Fig:numun}), we find
convergence as for $E_{1.5}$.  Like before, $E_3$ depends linearly on
the physical viscosity,
\begin{equation}
  \label{Gl:nunum-3}
  E_{3} \approx 0.96 \frac{\nu}{10^{-5}}.
\end{equation}
The inner vortex instabilities present at all but the finest grid
complicate the interpretation of the results.  They lead to a
comparatively strong difference between the late stages of the runs with
$n \le 2048$ and the reference run, which is reflected in the upturn
of the black curve in the right panel of \figref{Fig:numun}.
Simulations with the generally more viscous PLM schemes, on the other
hand, suppress inner-core instabilities and, to a lesser degree, also
outer-filament instabilities.  Since the absence of inner-vortex
instabilities renders the series of PLM runs close to the reference run,
the total error, $E_{3}$, the difference between their numerical
viscosity and that of the MP5 runs is relatively low.

\section{Summary and conclusions}
\label{Sek:Concl}

We estimated the numerical viscosities of simulations of the
non-magnetic KHI in two spatial dimensions based on several series of
simulations varying the grid resolution by a factor of 32.  In a first
step, we determined the dependence of the amplitude of the mode
excited by the initial perturbations on the physical viscosity.  This
result was then applied to assess the effective numerical viscosity of
simulations in ideal HD.  This procedure works very well in the early
phase of the evolution, during which the KHI grows exponentially.  We
find a power-law scaling of the numerical viscosity on the grid
resolution with an exponent slightly above and slightly below 2 for
runs with high-order MP5 and second-order PLM reconstruction,
respectively.  After the termination of the growth, however, the
dynamics is more complex, involving the appearance of secondary
instabilities on the KH vortices.  Their growth rates depend strongly
on the resolution and differ between reconstruction methods, causing
the same analysis to yield worse results than in the earlier phase.
We note that The ansatz of \eqref{Gl:nu*} is only well motivated in 
the linear regime.  Its applicability in the saturated phase requires
additional study.

This relative failing of the method hints towards the necessity of
applying the same rigorous analysis used for, e.g., the tearing-mode
instability by
\cite{Rembiasz_et_al__2017__apjs__OntheMeasurementsofNumericalViscosityandResistivityinEulerianMHDCodes}.
We suspect that, instead of determining the global
diffusive/dissipative error of a model, a breakdown of their
dependence in terms of the characteristic velocity and length scales
would serve to explain the nature of numerical viscosity.  Such an
analysis would be particularly helpful for the development or
suppression of the secondary instabilities, which grow on much shorter
length scales as the primary KHI.  While this kind of study would thus
be worthwhile for the non-linear saturated state, it is
beyond the scope of this article.  Further relevant extensions might
focus on the effect of magnetic fields and three-dimensional systems.

\section*{Acknowledgements}
\label{Ackno}

MO acknowledges support from the European Research Council (ERC; FP7)
under ERC Starting Grant EUROPIUM-677912 and from the the Deutsche
Forschungsgemeinschaft (DFG, German Research Foundation) --
Projektnummer 279384907 -- SFB 1245.  MAA acknowledges support from
the Spanish Ministry of Economy and Competitiveness (MINECO) through
grants AYA2015-66899-C2-1-P, and MTM2014-56218- C2-2-P, and from the
Generalitat Valenciana (PROMETEOII-2014- 069, ACIF/2015/216).

\section*{Bibliography}
\label{Sek:Bib}

\providecommand{\newblock}{}

\end{document}